\documentclass[preprintnumbers,amsmath,amssymbm,preprint]{revtex4}
\usepackage{epsfig}
\usepackage{graphicx}

\begin{document}
\title{Analytic toy-model for the ISCO shift}
\author{Shahar Hod}
\address{The Ruppin Academic Center, Emeq Hefer 40250, Israel}
\address{ }
\address{The Hadassah Institute, Jerusalem 91010, Israel}
\date{\today}

\begin{abstract}
A simple black-hole-ring system is proposed as a toy model for the
two-body problem in general relativity. This toy-model yields the
fractional shift
$\Delta\Omega_{\text{isco}}/\Omega_{\text{isco}}={{29}\over{81\sqrt{2}}}\eta$
in the Schwarzschild ISCO (innermost stable circular orbit)
frequency, where $\eta\equiv m/M_{\text{ir}}\ll 1$ is the
dimensionless ratio between the mass of the particle and the
irreducible mass of the black hole.
%This {\it analytical} expression
%is astonishingly close ($\sim 1\%$ difference) to the {\it
%numerically} computed value.
Our model
%reveals the underlying mechanism responsible for the ISCO
%shift phenomena. It
suggests that the second-order spin-orbit
interaction between the black hole and the orbiting particle (the
dragging of inertial frames) is the main element determining the
observed value of the ISCO shift.
\end{abstract}
%\bigskip
\maketitle
%]

{\bf Introduction.} The geodesic orbits of test particles in
black-hole spacetimes have been studied extensively by many
researches, see \cite{Bar,Chan,Shap,CarC,Fav} and references
therein. The innermost stable circular orbit (ISCO) is especially
interesting from both an astrophysical and theoretical points of
view. This orbit is defined by the onset of a dynamical instability
for circular geodesics -- it separates stable circular orbits from
orbits that plunge into the central black hole \cite{Chan}. This
special orbit is therefore important in the context of inspiralling
compact binaries since it represents the critical point where the
character of the orbit (and thus also the character of the
corresponding emitted gravitational radiation) sharply changes
\cite{Fav}. In addition, this marginally stable orbit is usually
regarded as the inner edge of accretion disks around central black
holes \cite{Chan}.

An important physical quantity which characterizes the ISCO is the
orbital angular frequency $\Omega_{\text{isco}}$ as measured by
asymptotic observers. This gauge-invariant frequency is often
regarded as the end point of the inspiral gravitational templates
\cite{Fav}. For a test-particle in the Schwarzschild spacetime, this
frequency is given by $M_{\text{ir}}\Omega_{\text{isco}}=6^{-3/2}$
\cite{Bar,Chan,Shap,CarC,Fav}, where $M_{\text{ir}}$ is the
irreducible mass \cite{Noteirr} of the central black hole.

What happens to the ISCO frequency if the mass $m$ of the particle
can no longer be neglected as compared to the mass $M_{\text{ir}}$
of the black hole? In this case one should take into account the
gravitational self-force (GSF) corrections to the orbit
\cite{Ori,Poi,Lou1,Det1,Bar1,Det2,Sag,Kei,Sha,Dam,Bar2,Fav2}. The
gravitational self-force has two distinct contributions: (1) It is
responsible for dissipative effects that cause the orbiting particle
to lose energy and angular momentum to gravitational waves. The
location of the ISCO may become blurred due to these
non-conservative effects \cite{Ori,Fav}. (2) The gravitational
self-interaction is also responsible for conservative effects which
preserve the characteristic constants of the orbital motion. This
effect is responsible for a non-trivial shift in the orbital
frequency of the ISCO.

Calculations of the GSF effects are mainly motivated by the need to
analyze in detail extreme-mass-ratio inspirals in which a compact
object interacts with a massive black hole \cite{Fav}. These
two-body events with $m/M_{\text{bh}}\lesssim 10^{-3}$ are expected
to be an important source for the Laser Interferometer Space Antenna
(LISA) \cite{Lisa}.

The computation of the GSF correction to the orbit is a highly
non-trivial task. After a decade of intensive efforts by many groups
of researches to evaluate the effects of the self-interaction terms
on the orbital parameters, Barack and Sago \cite{Bar2} have recently
succeeded in calculating the shift in the ISCO frequency due to the
conservative part of the GSF in the Schwarzschild spacetime. Their
numerical result for the ISCO frequency shift can be expressed in
the form \cite{Dam,Bar2,Notecon}:
\begin{equation}\label{Eq1}
M\Omega=6^{-3/2}[1+c\cdot \eta+O(\eta^2)]\ \ \ \text{with}\ \ \
c\simeq 1.251\ ,
\end{equation}
where $M\equiv M_{\text{ir}}+m$ is the total mass and $\eta\equiv
m/M_{\text{ir}}$ is the dimensionless ratio between the masses. The
result (\ref{Eq1}) is especially important because it provides
gauge-invariant information about the strong-gravity effects in the
vicinity of the black hole.

The main goal of the present study is to analyze a simple toy model
which captures some of the essential features of the (physically
more interesting) two-body problem in general relativity. In
particular, we would like to provide a simple and intuitive
explanation for the {\it increase} in the ISCO frequency in the
extreme-mass-ratio limit [see Eq. (\ref{Eq1})]. The proposed toy
model is composed of a stationary axisymmetric ring of particles in
orbit around a black hole. As shown by Will \cite{Will}, this
composed system is amenable to a perturbative analytic treatment.

{\bf The toy-model.} It is well-known \cite{Will} that local
inertial frames are {\it dragged} by an orbiting particle. In fact,
because of the dragging of inertial frames by the orbiting particle,
one can have a Schwarzschild-like black hole with zero angular
momentum but with a {\it non}-zero angular velocity, see Eq.
(\ref{Eq6}) below.

We propose to model the conservative behavior of the
black-hole-particle system using the analytically solvable model of
the composed black-hole-ring system. We expect this toy model to
capture the essential features of the original black-hole-particle
system, at least qualitatively. In particular, like the orbiting
particle, the rotating ring can drag the generators of the
black-hole horizon \cite{Will}. We therefore expect the
(analytically-known) spin-orbit interaction between the black hole
and the ring to mimic, at least qualitatively, the corresponding
spin-orbit interaction in the original black-hole-particle system.

To analyze the influence of the frame-dragging effect (caused by the
azimuthal motion of the ring) on the ISCO frequency, we must focus
on the {\it second-order} spin-orbit interaction \cite{Notesp}
between the ring and the black hole. This interaction introduces
terms of order $O[(mj)^2/M^3_{\text{ir}}]$ into the energy budget of
the system [see Eqs. (\ref{Eq2}) and (\ref{Eq6}) below] and an
analogous term of order $O(mj/M^3_{\text{ir}})$ to the expression of
the orbital frequency [see Eq. (\ref{Eq9}) below], where $mj$ is the
orbital angular momentum of the ring.

The black-hole-ring system, which is composed of a stationary
axisymmetric ring of particles in orbit around a slowly rotating
black hole, was analyzed by Will \cite{Will}. This system is
characterized by five physical parameters \cite{Will}: The
irreducible mass $M_{\text{ir}}$ of the black hole, the angular
velocity $\omega_{\text{H}}$ of the horizon, the rest mass $m$ of
the ring, the proper circumferential radius $R$ of the ring, and the
half-thickness $r\ll R$ of the ring.

To second order in the mass $m$ of the ring and to second order in
the angular velocity $\omega_{\text{H}}$ of the black-hole horizon,
a sequence of black-hole-ring equilibrium configurations was
obtained in \cite{Will}. The total gravitational mass of the
composed black-hole-ring configuration is given by \cite{Will}
\begin{eqnarray}\label{Eq2}
E(x;M_{\text{ir}},\omega_{\text{H}},j)=M_{\text{ir}}+
2M^3_{\text{ir}}\omega^2_{\text{H}}+m-m\Phi(x)-\omega_{\text{H}}
mj\Psi(x)-{{m^2x}\over{2\pi
M_{\text{ir}}}}\ln\Big({{8M_{\text{ir}}\over{xr}}}\Big)\  ,
\end{eqnarray}
where
\begin{equation}\label{Eq3}
x\equiv M_{\text{ir}}/R\ \ \ ; \ \ \ \Phi(x)\equiv
1-{{1-2x}\over{(1-3x)^{1/2}}}\ \ \ ; \ \ \ \Psi(x)\equiv
12{{x^3-2x^4}\over{1-3x}}\  ,
\end{equation}
and $j$ is the angular momentum per unit mass of the ring which is
given (to the necessary order) by \cite{Will}
\begin{equation}\label{Eq4}
j={{M_{\text{ir}}}\over{[x(1-3x)]^{1/2}}}\  .
\end{equation}

Each term on the r.h.s. of Eq. (\ref{Eq2}) has a clear physical
interpretation \cite{Will}:
\begin{itemize}
\item{The first two terms on the r.h.s. of (\ref{Eq2}) represent the contribution of the black hole to the total
mass of the system. Remembering that a slowly rotating vacuum Kerr
black hole is characterized by the relation
$M_{\text{Kerr}}=M_{\text{ir}}+2M^3_{\text{ir}}\omega^2_{\text{H}}+O(\omega^4_{\text{H}})$,
one can identify the second term of (\ref{Eq2}) as the
rotational-energy of the spinning black hole. Note, however, that
for the black-hole-ring system $\omega_{\text{H}}$ contains a term
linear in $mj$ [see Eq. (\ref{Eq6}) below], and thus the rotational
energy of the black hole contains within it a second-order
self-interaction term of order $O[(mj)^2]$.}
\item{The third term on the r.h.s. of (\ref{Eq2}) is simply the ``bare"
(rest)
mass of the ring.}
\item{The fourth term on the r.h.s. of (\ref{Eq2}) represents the
first-order mutual interaction  between the ring and the black hole.
In the large-$R$ (small-$x$) limit it becomes $-M_{\text{ir}}m/2R$,
which can be identified as the negative Newtonian potential energy
of the black-hole-ring system plus the Newtonian rotational energy
of the ring.}
\item{The fifth term on the r.h.s. of Eq. (\ref{Eq2}) represents a
spin-orbit interaction between the spinning black hole (which is
characterized by the horizon angular velocity $\omega_{\text{H}}$)
and the rotating ring (which is characterized by the angular
momentum $mj$). We expect this term to capture, at least
qualitatively, the essential features of a similar spin-orbit
interaction in the original black-hole-particle system. Since
$\omega_{\text{H}}$ contains a term linear in $mj$ [see Eq.
(\ref{Eq6}) below], this spin-orbit interaction term contains within
it a second-order self-interaction term of order $O[(mj)^2]$.}
\item{The sixth term on the r.h.s. of Eq. (\ref{Eq2}) represents the
second-order gravitational self-energy of the ring \cite{Tho} (not
considered in \cite{Will}).
%where $r\ll R$ is the ring half-thickness.
This term reflects the inner interactions between the {\it many}
particles that compose the ring. Since our main interest here is the
original two-body system with a {\it single} orbiting particle (we
only use the analytically-known properties of the rotating ring to
model the rotation of the particle), we shall not consider this
many-particle term [and an analogous term in Eq. (\ref{Eq9}) below]
here. This will allow us to focus on the influence of the {\it
frame-dragging} effect {\it alone} on the ISCO frequency. In this
respect, the ring considered in \cite{Will} should be regarded as a
quasi test ring.}
\end{itemize}

The total angular momentum of the composed black-hole-ring system is
given by \cite{Will}
\begin{equation}\label{Eq5}
J(x;M_{\text{ir}},\omega_{\text{H}},j)=4M^3_{\text{ir}}\omega_{\text{H}}-8mjx^3+mj\
.
\end{equation}
The first two terms on the r.h.s. of (\ref{Eq5}) represent the
angular momentum $J_{\text{H}}$ of the black hole while the third
term is the angular momentum of the ring \cite{Will}. As emphasized
in \cite{Will}, the simple relation
$\omega_{\text{H}}=J_{\text{Kerr}}/4M_{\text{bh}}M^2_{\text{ir}}$
between the angular momentum of the black hole and the angular
velocity of the horizon, which was valid for {\it vacuum} Kerr black
holes [and, in particular, the simple relation
$\omega_{\text{H}}(J_{\text{Kerr}}=0)=0$] no longer holds when
matter (the ring) is present in the black-hole exterior. In
particular, due to the dragging of inertial frames \cite{Will} by
the rotating ring, a {\it zero} angular momentum black hole
($J_{\text{H}}=0$) is characterized by a {\it non}-zero angular
velocity [see Eq. (\ref{Eq5})]:
\begin{equation}\label{Eq6}
\omega_{\text{H}}={{2x^3}\over{M^3_{\text{ir}}}}\cdot mj\  .
\end{equation}

Substituting Eqs. (\ref{Eq4}) and (\ref{Eq6}) into Eq. (\ref{Eq2}),
one finds
\begin{eqnarray}\label{Eq7}
E(x;M_{\text{ir}},\omega_{\text{H}},j)=M_{\text{ir}}+{{1-2x}\over{(1-3x)^{1/2}}}\cdot
m+ {{8x^5(-2+3x)}\over{(1-3x)^2}}\cdot{{m^2}\over{M_{\text{ir}}}}\
\end{eqnarray}
for the total energy of the composed black-hole-ring system.

{\bf The ISCO.} A standard way to identify the location of the ISCO
is by finding the minimum of the total energy
\cite{Fav,Fav2,Bon,Noteisc}. A simple derivation of (\ref{Eq7}) with
respect to $x$ reveals that the perturbed ISCO is characterized by
\begin{equation}\label{Eq8}
x_{\text{isco}}={1\over 6}+{{5\sqrt{2}}\over{243}}\eta+O(\eta^2)\  .
\end{equation}
Substituting (\ref{Eq8}) into the expression \cite{Will}
\begin{equation}\label{Eq9}
M_{\text{ir}}\Omega=x^{3/2}-4M_{\text{ir}}\omega_{\text{H}}x^3+O(\omega^2_{\text{H}})\
\end{equation}
for the angular velocity of the rotating ring, one finds
\begin{equation}\label{Eq10}
M\Omega_{\text{isco}}=6^{-3/2}\Big[1+\Big(1+{{29}\over{81\sqrt{2}}}\Big)\eta+O(\eta^2)\Big]\
\end{equation}
for the perturbed ISCO frequency.

{\bf Summary.} A black-hole-ring system was proposed as a simple toy
model for the (physically more interesting) two-body problem in
general relativity. In particular, we have used the ({\it
analytically-known}) second-order spin-orbit coupling between the
black hole and the orbiting ring to describe the essential features
of a similar spin-orbit interaction in the original
black-hole-particle system. Our original motivation was to provide a
simple {\it qualitative} explanation for the increase in the ISCO
frequency in the extreme-mass-ratio limit. Admittedly, we had no a
priori reason to believe that this simple toy model would be able to
provide a good {\it quantitative} description of the ISCO shift
phenomena of the original black-hole-particle system. However,
somewhat surprisingly, our {\it analytical} expression
$c=1+{{29}\over{81\sqrt{2}}}\simeq 1.253$ for the value of the ISCO
shift-parameter
%for the fractional shift in the ISCO frequency,
%$\Delta\Omega_{\text{isco}}/\Omega_{\text{isco}}={{29}\over{81\sqrt{2}}}\eta\simeq
%0.253\eta$,
is astonishingly close
%($\sim 0.2\%$ difference)
to the {\it numerically} computed \cite{Bar2,Akc} value
%$\Delta\Omega_{\text{isco}}/\Omega_{\text{isco}}
$c\simeq 1.251$. This fact suggests that the second-order spin-orbit
interaction between the black hole and the orbiting object (the
dragging of inertial frames) is the main element determining the
observed value of the ISCO shift.

\bigskip
\noindent
{\bf ACKNOWLEDGMENTS}
\bigskip

This research is supported by the Carmel Science Foundation. I thank
Yael Oren, Arbel M. Ongo and Ayelet B. Lata for helpful discussions.

%\newpage


\begin{thebibliography}{99}
\bibitem{Noteirr} The irreducible mass is closely related to the black-hole surface
area: $M_{\text{ir}}=({\text{area of horizon}}/16\pi)^{1/2}$. In the
Schwarzschild case the irreducible mass coincides with the mass of
the black hole, $M_{\text{ir}}=M_{\text{bh}}$.
\bibitem{Notecon} As emphasized in \cite{Bar2}, the work \cite{Bar2} is
complementary to \cite{Ori} in that \cite{Bar2} considers only
conservative effects and ignore dissipation. It is only then that
the ISCO has a sharp location.
\bibitem{Notesp} Note that this interaction is second-order in the
orbital angular momentum.
\bibitem{Noteisc} The critical orbit so defined is sometimes referred to as the MECO
(maximum binding-energy circular orbit) \cite{Fav,Fav2,Bon}.
\end{thebibliography}
\end{document}